\documentstyle[12pt,aasms4]{article}

\def\listitem{\par \hangindent=50pt\hangafter=1
     $\ $\hbox to 20pt{\hfil $\bullet$ \hfil}}

\def\puncspace{\ifmmode\,\else{\ifcat.\C{\if.\C\else\if,\C\else\if?\C\else%
\if:\C\else\if;\C\else\if-\C\else\if)\C\else\if/\C\else\if]\C\else\if'\C%
\else\space\fi\fi\fi\fi\fi\fi\fi\fi\fi\fi}%
\else\if\empty\C\else\if\space\C\else\space\fi\fi\fi}\fi}
\def\SP{\let\\=\empty\futurelet\C\puncspace}

\def\kms{km~s$^{-1}$\SP}

\def\degree{$^\circ$\SP}
\def\h1{$h^{-1}$\SP}

\def\etal{{\rm et al.\/}\ }

\def\apj{ApJ,}

\def\mnras{MNRAS,}

\def\aj{AJ,}
\def\lsim{~\rlap{$<$}{\lower 1.0ex\hbox{$\sim$}}}
\def\gsim{~\rlap{$>$}{\lower 1.0ex\hbox{$\sim$}}}
\def\void#1{{}}

\parskip=\baselineskip
\newcommand{\mincir}{\raise -2.truept\hbox{\rlap{\hbox{$\sim$}}\raise5.truept
\hbox{$<$}\ }}
\newcommand{\magcir}{\raise -2.truept\hbox{\rlap{\hbox{$\sim$}}\raise5.truept
\hbox{$>$}\ }}
\newcommand{\minmag}{\raise-2.truept\hbox{\rlap{\hbox{$<$}}\raise 6.truept\hbox
{$>$}\ }}
\def\be{\begin{equation}}
\def\ee{\end{equation}}

\newcommand{\ba}{\begin{eqnarray}}
\newcommand{\ea}{\end{eqnarray}}
\newcommand{\brr}{\begin{array}}
\newcommand{\err}{\end{array}}
\newcommand{\bc}{\begin{center}}
\newcommand{\ec}{\end{center}}

\newcommand{\et}{{\em et al.}\,\,}

\def\listitem{\par \hangindent=50pt\hangafter=1
     $\ $\hbox to 20pt{\hfil $\ubfllet$ \hfil}}
\def\puncspace{\ifmmode\,\else{\ifcat.\C{\if.\C\else\if,\C\else\if?\C\else%
\if:\C\else\if;\C\else\if-\C\else\if)\C\else\if/\C\else\if]\C\else\if'\C%

\else\space\fi\fi\fi\fi\fi\fi\fi\fi\fi\fi}%
\else\if\empty\C\else\if\space\C\else\space\fi\fi\fi}\fi}
\def\SP{\let\\=\empty\futurelet\C\puncspace}

\def\kms{km~s$^{-1}$\SP}

\def\degree{$^\circ$\SP}
\def\h1{$h^{-1}$\SP}

\def\etal{{\it et al.\/}\ }

\begin{document}

\title{Redshifts for 2410 Galaxies in the Century Survey Region
\altaffilmark{1,2}}

\author{Gary Wegner and John R. Thorstensen}
\affil{Department of Physics \& Astronomy, 6127 Wilder Laboratory,
Dartmouth College, Hanover, NH 03755-3528}

\author{Michael J. Kurtz, Warren R. Brown, Daniel G. Fabricant, Margaret J.
Geller}
\author {John P. Huchra}  

\affil{Smithsonian Astrophysical Observatory, 60 Garden Street,
Cambridge, MA 02138}
\author {Ronald O. Marzke}
\affil {Department of Astronomy and Physics, San Francisco State University,
San Francisco, CA}
\author{Shoko Sakai} 
\affil{Division of Astronomy and Astrophysics, UCLA, Los Angeles, CA 90095}

\altaffiltext{1} {Work reported here based partly on
observations obtained at the Michigan-Dartmouth-MIT Observatory}
\altaffiltext{2} {Work reported here based partly on observations at the
Multiple Mirror Telescope, a joint facility of the Smithsonian Institution
and the University of Arizona}

\begin{abstract}

The `Century Survey'  strip
covers 102 square degrees within the limits 
$8.5^h \leq \alpha_{1950} \leq 16.5^h$, $29.0^\circ \leq \delta_{1950}
\leq 30.0^\circ$. The strip passes 
through the Corona Borealis supercluster and the outer
region of the Coma cluster. 

Within the Century Survey region, we have measured 2410 redshifts 
which constitute four
overlapping complete redshift surveys: (1) 1728 galaxies with
Kron-Cousins $R_{phot}  \leq 16.13$ covering the entire strip,
(2) 507 galaxies with $R_{phot} \leq 16.4$ in the
right ascension range $8^h 32^m \leq \alpha_{1950} \leq 10^h 45^m$,
(3) 1251 galaxies with
absorption- and K-corrected $R_{CCD, corr} \leq 16.2$ 
covering the right ascension range $8.5^h \leq \alpha_{1950} \leq 13.5^h$
and (4) 1255 galaxies with
absorption- and K-corrected $V_{CCD, corr} \leq 16.7$ also covering
the right ascension range $8.5^h \leq \alpha_{1950} \leq 13.5^h$.
All of these redshift samples are more than 98\% complete to the
specified magnitude limit.

We derived samples (1) and (2) from scans of the POSS1 red (E) plates calibrated
with CCD photometry. We derived samples (3) and (4) from  
deep V and R CCD images covering the entire region.

We include coarse morphological types for all of the galaxies in 
sample (1). The distribution of $(V - R)_{CCD}$ for each type
corresponds appropriately with the classification.   

\end{abstract}


\section{Introduction}

Redshift surveys have come of age. A host of samples derived from 
optical imaging surveys cover sizable solid angles and contain 
more than 1000 galaxies
(\cite{Davis82}; \cite{gel89}; \cite{gio89}; \cite{loveday92};
\cite{dac94}; \cite{ratcliffe96}; \cite{she96};
\cite {vet97};\cite{york00}; \cite{cross01}). 
The surveys range from recent mega-projects like the 
2dF(\cite{cross01}) and Sloan (\cite{york00}) surveys to 
the smaller ESO Key Program (\cite{vet97}) and the
Century Surveys  which we are discussing here (Geller \etal 1997).
Only three of these surveys are in the R-band: the Las Campanas 
(\cite{she96}), Sloan (\cite{york00})
and Century (\cite{ocs}) surveys.

We acquired the 2410 redshifts in this paper
to complete four redshift surveys in
the Century Survey strip:
(1) the original Century Survey of 1728 galaxies 
Kron-Cousins (Kron, White, \& Gascoigne 1953) $R_{phot} \leq 16.13$ covering the entire strip
(Geller \etal 1997; OCS hereafter),
(2) a survey including 508 galaxies with $R_{phot} \leq 16.4$ in the
right ascension range $8^h 32^m \leq \alpha_{1950} \leq 10^h 45^m$
(DCS hereafter),
(3) a survey of 1251 galaxies with
absorption- and K-corrected $R_{CCD, corr} \leq 16.2$ 
covering the right ascension range $8.5^h \leq \alpha_{1950} \leq 13.5^h$
(Brown \etal 2001; RCS hereafter)
and (4) a survey of 1255 galaxies with
absorption- and K-corrected $V_{CCD, corr} \leq 16.7$ also covering
the right ascension range $8.5^h \leq \alpha_{1950} \leq 13.5^h$
(Brown \etal; VCS hereafter).
All of these sets of redshifts are more than 98\% complete to the
specified magnitude limit.

The OCS  cuts
through the Great Wall and provides a sample of 
galaxies 2.3 magnitudes fainter than the characteristic $L_*$ magnitude
within it. The redshift survey also
includes the Corona Borealis supercluster ($\alpha_{1950} = 15.3^h$
to $15.6^h; \delta_{1950}= 27.5$ to $ 32$\degree) 
containing the galaxy cluster A2079 and the outer fringes of 
A1656, the
Coma cluster ($\alpha_{1950}=12^h57.5^m; \delta_{1950}= $28\degree15\arcmin).
In addition, the survey samples the Abell galaxy
clusters A690, A1185, A1213, A2162, and A2175.

We include the photographic magnitudes, $R_{phot}$, used to define
the OCS and DCS samples. For the galaxies in the OCS, we list morphological
types. To our knowledge, there are no other complete redshift surveys to
this depth with morphological types.

Section 2 discusses the photographic photometry. We use the CCD
photometry for the RCS and VCS (Brown \etal 2001) to examine residual
systematics in the photographic photometry. We also use the
CCD photometry to show that the morphological types
are sensible.  Section 3 
describes the spectroscopic observations and data reduction procedures.  
Section 4 gives the catalog of redshifts, photographic magnitudes, and
morphological types along with a brief discussion of the OCS and DCS
samples derived from the photographic photometry.
We conclude  in Section 5.

\section{Photometric Data and Sample Selection}

We constructed the OCS galaxy catalog from scans of the POSS1 E plates
according to the procedures outlined
by Kurtz et al. (1985).

For each galaxy in the catalog 
we derived an isophotal magnitude to
a  bright limiting isophote which varies unavoidably from plate to plate.
Two drift
scans from $8^{\rm h}27^{\rm m}$ to $11^{\rm h}55^{\rm m}55^{\rm s}$
and from $11^{\rm h}50^{\rm m}$ to $15^{\rm h}45^{\rm m}$ provided
the basis for the magnitude 
calibration
(Ramella \etal 1995; Kent \etal 1993).
The drift scans are both centered at $\delta = 29.5^\circ$.
The drift scan for early $\alpha$'s was done with the 1.2-m
telescope and for late
$\alpha$'s with the 61-cm telescope (now retired) of F. L. Whipple Observatory
(FLWO). We also used pointed observations to check the drift scans and
to calibrate the three POSS
plates E924, E1365, and E134 which cover the right ascension ranges
$8^{\rm h}32^{\rm m}32^{\rm s}$
to $8^{\rm h}58^{\rm m}50^{\rm s}$ and
$15^{\rm h}53^{\rm m}$ to $16^{\rm h}19^{\rm m}44^{\rm s}$.

We used an iterative procedure to calibrate the photographic photometry.
First, we fixed a preliminary zero-point on each plate by comparing
the instrumental magnitudes with the drift scans of Ramella \etal (1995) and
Kent \etal (1993) at $R = 16.0$. We then combined data from all plates and
obtained a preliminary global slope for the scale error. We then 
redetermined the zero-point (at $R = 16.0$) for each plate and fit a new 
global slope. The procedure converged after two iterations.

Recently Brown \etal (2001)  obtained $V$ and $R_{KC}$ photometry
for 1295 galaxies in the Century strip from CCD images 
with median rms errors of
$\pm$0.042 mag. in both $R_{CCD}$ and $V_{CCD}$.
By comparing the galaxy coordinates from the CCD data with
those from the PDS scans of the POSS1 plates, we estimate the rms
error in the  photographic coordinates,
$\Delta \theta = \pm 0.29^{\prime\prime}$.

The extensive CCD photometry enables a clean evaluation of the
RMS error in the photographic magnitudes and of the residual systematic
errors.
Figure 1 compares the photographic CS photometry with 
the CCD photometry of Brown \etal (2001). The zero-point offset
$\Delta = R_{CCD} - R_{phot} = 0.014 \pm 0.22$, is well within the 
average zero-point error in the CCD photometry ($\pm 0.034$ mag).
The $\pm 0.22$ magnitude scatter of the photographic relative to the CCD
photometry is consistent with the $\pm 0.25$ magnitude photographic scatter
estimated by Geller et al. (1997).

\centerline { EDITOR: PLEASE PLACE FIGURE 1 HERE}

Figure 2 shows $\Delta$ as a function of $R_{CCD}$  for 935 CS galaxies.
The comparison indicates a residual slope of 0.07 mag/mag, a 10\% error in
the original zero-point slope. The brightest $R < 11$ CS galaxies 
are $\sim$ 0.5 magnitudes brighter than the photographic magnitude; the
offset decreases for fainter galaxies. This error is consistent with variation in
the non-linearity from plate to plate.

\centerline { EDITOR: PLEASE PLACE FIGURE 2 HERE}

Figure 3 shows $\Delta$ as a function of the peak R surface brightness
returned by Sextractor (Bertin \& Arnouts 1996)
from the CCD data for 429  galaxies with 15.5$\leq R_{CCD} \leq
16.13$. In this apparent magnitude range the zero-point offset
between the photographic and CCD magnitudes is small. The sample of galaxies
in Figure 3 have a cleanly defined surface brightness peak (i.e. there are
no cosmic rays or bad columns, no superimposed stars, etc.). For the lowest
surface brightness galaxies, the photographic magnitudes are brighter than
the CCD magnitudes; for high surface brightness galaxies, the effect is the
reverse. The sense of the systematic deviations is as expected from
the effect of saturation on the photographic magnitudes. 
The dashed line in Figure 3 shows the best-fit relation
$\Delta = 0.13 \mu_R -2.59$ where $\mu_R$ is the peak
R-band surface brightness. 

\centerline { EDITOR: PLEASE PLACE FIGURE 3 HERE}

Figure 4 shows $\Delta$ as a function of $(V - R)_{CCD}$ for the 
508 galaxies in the same range of apparent magnitude as in Figure 3.
The photographic magnitudes are too bright for the blue (generally low
surface brightness) galaxies  and too faint for the red (generally
higher surface brightness) galaxies. The best-fit
relation (dashed line) is $\Delta = -0.66(V - R)_{CCD} + 0.36.$
Because central surface brightness and color are correlated, the
underlying systematic effect is the same here as in Figure 3.

\centerline {EDITOR: PLEASE PLACE FIGURE 4 HERE}

This correlation is shown in Figure 5 which is a plot of surface
grightness, $\mu_R$, against color index $(V-R)_{CCD}$ for the same sample
of galaxies as in Figures 3 and 4. 

\centerline {EDITOR: PLEASE PLACE FIGURE 5 HERE}

Table 1 lists the 4 magnitude limited redshift surveys which can be extracted
from the data in Tables 2 and 3 (along with the photometry from Table 3 of
Brown et al. (2001)) and their properties.
In addition to the two samples,
the OCS and DCS, originally derived from the plate scans,
Brown \etal (2001) constructed two more rigorously defined magnitude
limited redshift surveys, the RCS and VCS, in the
64 square degree region with CCD photometry.
The uncorrected photographic photometry in Table 2 determines the
limiting magnitude for the OCS and DCS surveys. Constructing samples
from photometric surveys before correction for systematic effects has been
standard procedure. However, as emphasized Brown \etal (2001), 
making absorption and K-corrections before magnitude limiting the sample
ensures that the redshift survey samples galaxies of all spectroscopic
types to the same effective depth. We derived
the RCS and VSC samples {\it after} absorption- and K-correcting the
relevant CCD magnitudes. Table 1 lists both the total number of galaxies in
each sample and the number with redshifts; all of the samples are more than
98\% complete to the specified limit.

\centerline { EDITOR: PLEASE  PLACE TABLE 1 HERE}

\subsection {Morphological Types} 

One of us (GAW) used a 30 power
magnifier to  estimate morphological types for all OCS galaxies with 
photographic $R_{phot} \leq 16.13$ from the glass copies of the 
POSS1 O (blue) plates, 
which for this work were found to be superior to the
paper and digitized copies. With mounting distance the
classifications are increasingly difficult. At the largest redshifts
this amounts to an estimate of the disk-to-bulge ratios which can still be
seen when spiral structure is no longer resolved. Consequently the
Hubble types are in coarse
bins: E, S0, Sa, Sb, Sc, and Irr plus the corresponding barred
types. Double morphological type entries in Table 2 indicate galaxies
with images on more than one
plate; from these overlaps we estimate that
the classification error is $\pm$1 type. 

\centerline { EDITOR: PLEASE PLACE FIGURE 6 HERE}

For the region with CCD photometry,   
Figure 6 shows the $(V - R)_{CCD}$ color distribution for each 
morphological type. We
do not distinguish between barred and unbarred members of the class. 
From E through Sc, the mean color for
each morphological type corresponds remarkably well to
the colors quoted by Fukugita, Shimasaku \& Ichikawa (1995; FSI, Table 3a).
Although sparsely populated, our color
distribution for the Im/Irr class, appears bimodal. The color of the blue
peak corresponds to the typical V-R$_C$ color (0.31) quoted by
FSI. The CCD images reveal that many objects in the red ``peak'' are
tight pairs: some of these show obvious tidal distortions and others
are pairs of early-type galaxies. Most of these pairs are not clearly
resolved on the POSS1 plates.

\section{Spectroscopic Observations}

Tables 2 and 3 list a total of 2437 galaxies  in the CS region. 
All but 27 of the galaxies in Table 2 have redshifts;
all but 14 redshifts were measured with Dartmouth or CfA facilities
including the
2.4 m Hiltner telescope of the Michigan-Dartmouth-MIT (MDM) Observatory on 
Kitt Peak, the Multiple Mirror Telescope (MMT) on Mount Hopkins, 
and the 1.5 m telescope of the Fred Lawrence Whipple Observatory. 

We generally observed galaxies in the denser fields with the
2.4 m using the `Decaspec,' a ten-object fiber instrument  built for
this project. We used the MMT (in its original configuration) and the
Whipple Observatory 1.5 m to obtain individual 
spectra in  
more sparsely populated regions where the multiplexing feature of the 
Decaspec was not advantageous.

\subsection{MDM Decaspec Observations}

We measured 1019 redshifts with the Decaspec or the Mark III
Spectrograph on the 2.4m Hiltner Telescope at the 
MDM Observatory.
Fabricant \& Hertz (1990) describe the Decaspec,
a fiber-moving head mounted on the telescope in front of 
the spectrograph. The Decaspec has ten movable probes; each of the
probes contains 
five 2.3 arcsec diameter optical
fibers set in a row with 21 arcsec spacing. 
The probes run along parallel tracks over the 20 arcmin diameter field
of the 2.4 m Hiltner telescope. The motion of the probes along and
perpendicular to the tracks combined with instrument rotation enables
target acquisition with one fiber per probe.
The light from the Decaspec feeds into the Mark III spectrograph,
designed by W. A. Hiltner. 

We made all of the CS observations 
with a 300 lines/mm grism blazed at 5400 \AA.
The Decaspec observations began in 1989 May. We used a   
number of CCD detectors during the course of the observations.
All of these detectors delivered a resolution of $\sim$12 \AA ~FWHM and a spectral 
coverage of approximately 4000 - 7000 \AA. 

The observations usually consisted of three 40 minutes integrations
to enable cosmic ray rejection.  Our subsequent data
reductions to one-dimensional spectra used the `apextract' package in
IRAF{\footnote{IRAF is distributed by the National Optical Astronomy
Observatories which is operated by the Association of Universities for Research
in Astronomy, Inc. under contract with the National Science Foundation}} (Tody
1986). We used standard data reduction techniques
with custom scripts 
written to simplify the book keeping. We used 
the `imcombine' option to combine the three spectra and extracted the
resulting
spectrum with  `apsum.' We used continuum spectra
from a screen inside the dome for flattening; we constructed  
individual wavelength calibrations from  HgNeXe
comparison exposures before and after each object integration. Generally 
we fitted a
4th or 5th order polynomial to the non-linear part of the wavelength 
correction curve.

We used the four non-object fibers 
from each probe of the Decaspec for sky subtraction. We corrected differences
in the fiber transmission 
by normalizing the strong $\lambda$5577 night sky line equivalent
widths in all the spectra to the galaxy fiber and then subtracting the median
of the four spectra from the object spectrum. 
By dividing the
galaxy spectra by very high signal
to noise blue star spectra (with all stellar absorption lines
removed) normalized to 1.0
everywhere, we eliminated
the strong telluric night sky
absorptions in the red due to water and $\rm{O}_2$.

To produce template spectra, we observed velocity standards at least
once nightly. The templates 
included stars used routinely at the CfA along with
IAU radial velocity standards from the $Astronomical~Almanac.$ 
We obtained at least 10
separate spectra through one fiber in each of the 10 probes. We reduced these
spectra as described above; we then removed velocity shifts and summed the
spectra  to
produce a template.

Thorstensen (Thorstensen \etal 1989) wrote
the software to extract redshifts and their estimated errors. 
The software includes the
Tonry \& Davis (1979) 
cross-correlation algorithm for absorption line spectra and 
a multiple Gaussian fitting routine for emission lines.
We accepted only
cross-correlation velocities with high formal statistical confidence;
we checked all fits by eye.  For most objects, $R \gtrsim 3.0$
with $R > 10$  common.
The mean uncertainty in redshifts derived from the Decaspec
is $\pm44.5$ \kms.

We observed several hundred galaxies one-by-one with   
the Mark III spectrograph or the 1.5 FWLO spectrograph.
The data reductions were 
identical to those in the redshift survey of the fainter
Zwicky galaxies in the first CfA strip (Thorstensen \etal 1989, 1995, 
Wegner \etal 1990).

\subsection{MMT Observations}

We observed 1079 galaxies with the Multiple Mirror
Telescope (MMT). We used the blue channel of the MMT spectrograph and 
the photon-counting
Reticon system (Latham 1982) and, later, the blue and red channels with
CCD detectors for the observations. The 
reduction to heliocentric radial
velocity employed the RVSAO package of Kurtz \etal (1992) and Mink \& Wyatt 
1992) developed at the CfA, and closely resembles the MDM
procedures.
These techniques  are described
by Huchra \etal (1995, 1999). The typical external error in the MMT 
velocities
is $\sim35$\ \kms.

\subsection{Recent 1.5-meter FLWO Observations}

To complete the RCS and VCS (Brown et al. 2001), we measured
226 new redshifts with the FAST (Fabricant \etal 1998) spectrograph on the
Fred Lawrence Whipple Observatory 1.5m telescope in 2000 April and
2000 November; Table 3 lists these redshifts. From 1994 through 1996 we also used
FAST to measure 72 redshifts in the OCS (included in Table 2).  We used the 300 line/mm grating with 6 \AA\ 
resolution, and measured velocities with the cross-correlation
package RVSAO (Kurtz \& Mink 1998).  The mean uncertainty of the
velocities is $\pm40$ km s$^{-1}$.  

\subsection{Previously Published Redshifts}

Some of our redshifts 
have been published previously: 164 redshifts
are from Huchra \etal (1990), 92 are from
Thorstensen \etal (1989), and
11 are from Willmer \etal (1996).
The Huchra \etal (1990), Thorstensen \etal.
(1989) redshifts
were acquired using the same instruments and reduction
methods described above; they have the same zero points and
errors.

\section{The Catalog of Redshifts}

Table 2 lists the data for the OCS, for the DCS and for some fainter objects. 
Columns 1 and 2 are the epoch 2000.0, Right Ascension and Declination measured 
from the POSS1 plates (these coordinates agree very well with the
ones determined from the CCD images). Column 3 is the
$R_{phot}$ magnitude calibrated as in Geller \etal (1997) and as
described in Section 2.

Columns 4 and 5 contain the heliocentric redshift
and its error, respectively.
Column 6 is the estimated morphological type.
Column 7 contains notes
for objects with poor positions: IP indicates an interacting
galaxy pair, CG indicates
a companion galaxy, and NS indicates a nearby star.
Column 8 contains the sample designation (OCS or DCS). Column 9 contains GAWs
classification notes.

\centerline {EDITOR: PLEASE PLACE TABLE 2 HERE}

For galaxies required to complete 
the samples of Brown et al (2001), Table 3 lists epoch 2000.0
Right Ascension and Declination (columns 1 and 2, respectively),
columns 3 and 4 are the
heliocentric redshift and its error, respectively. 
Brown \etal (2001; Table 3) publish  CCD 
$V$ and $R_{KC}$ magnitudes, galactic extinctions, and K-terms  for {\it all}
of the galaxies in  their samples.

\centerline {EDITOR: PLEASE PLACE TABLE 3 HERE}

The cone diagrams in Figures 7 and 8 show the two complete redshift surveys
defined from the photographic photometry in Table 2. Figure
9 shows the corresponding redshift histograms. 
The Great Wall
contributes the peak at redshifts between 0.02 and 0.035.

Geller
\etal (1997) discuss the OCS ; they derive an R-band luminosity function
and luminosity density $j = 2.8 \pm 0.9 \times 10^8$L$_\odot$,
in excellent agreement with the value recently reported
by Blanton \etal (2001) who analyze a portion of the Sloan Digital Sky Survey.
Brown \et al (2001) derive an R-band luminosity function and 
corresponding luminosity density for the CCD-based RCS. Their results
are consistent with the photographic determination.
Brown \etal show further that the OCS omits few LSB galaxies with
$\mu(0)_R > 20.8$ mag arcsec$^{-2}$; the photographic catalog contains
12 LSB galaxies with $R_{phot} < 16.13$ and the CCD catalog contains 15 in
the overlap region.

\centerline {EDITOR: PLEASE PLACE FIGURES 7, 8 AND 9 HERE}

The DCS in Figure 8 is 0.27 magnitudes deeper than the OCS over the
right ascension range $8.5^h \leq \alpha_{1950} \leq 13.5^h$. 
There are 518 galaxies in this sample;
177 of them have $R_{phot} > 16.13$. Generally, the fainter galaxies
(open circles) populate
structures already defined by the brighter sample (filled squares). The
histogram in Figure 9b shows the redshift distribution for 
the  $R_{phot} < 16.13$ galaxies and for the fainter sample. It is 
again apparent that both sets of galaxies trace the same structures; the 
distribution of the fainter sample shifts toward
larger redshift as expected.

\section{Conclusions}

The 2410 redshifts we report here constitute four complete redshift
surveys in the Century Survey region.  We include
photographic photometry  and morphological types for most of the
galaxies. 

Papers in the literature discuss three of the surveys, the OCS, the RCS, and
the VCS. Brown \etal (2001) tabulate the V and R CCD photometry for the 
RCS and VCS. They also list absorption and K-corrections for the samples.

The DCS appears only in this paper. Not surprisingly, the fainter galaxies
in the DCS populate features defined by the original Century Survey 
in the region. 

Redshifts in the Century Survey region are useful for the analysis and interpretation
of other surveys. For example,
the depth of the OCS, RCS and VCS is comparable with the
depth of the 2MASS J-band survey (\cite{jarrett00}) and should provide a 
sizable 
magnitude limited sample for computation of the J-band luminosity
function and for some assessment of its morphological type dependence.

The surveys we discuss also cover 90\% of the KPNO International 
Spectroscopic Survey
(KISS) region (\cite{salzer00}). KISS is an objective-prism survey. 
Availability of the CS data enables comparison of the distribution
of the KISS sample galaxies with  those in a complete magnitude limited
redshift survey.

We thank Tim Beers for suggesting the `Century Survey' name.  We are grateful to Bill van Altena
for making the Yale PDS available to us and for making our
many plate-scanning visits to Yale pleasant. The Yale PDS facility 
was NSF-supported.  
At Dartmouth, this project was partially funded from Dartmouth College and
by two  funding agencies. 
JRT was supported in part by NSF Grant No. AST86-20081 and a 
Research Corporation Cottrell Grant. GW was supported in part by NSF Grants No.
AST86-20081, AST90-23450, and AST93-47714. We thank the MDM Observatory 
staff, especially Bob Barr for excellent support at the 2.4 m. 
The Smithsonian Institution funded this project at the CfA. 
We thank Emilio Falco and Rudy Schild for making some of the photometric
observations we used to calibrate the photographic photometry. 
The MMT night assistants, Carol Heller, John McAfee, and  Janet Miller,
provided invaluable support and Suzan Tokarz helped with data reductions.

\clearpage
\noindent FIGURE CAPTIONS
\vskip 0.2in
\figcaption [] {Histogram of $\Delta = R_{CCD} - R_{phot}$. 
The dotted line shows the zero-point offset, $\Delta = 0.014$ mag. The dispersion is $\pm$0.22
magnitudes, comparable with the $\pm 0.25$ magnitude error in $R_{phot}$
originally estimated by Geller \etal (1997).}

\figcaption []{Comparison of $R_{phot}$ with R$_{CCD}$. The dotted
line shows the zero-point offset $\Delta = R_{CCD} - R_{phot} = 0.014 \pm
0.22$ mag. The long dashed line shows the best-fit relation between
$\Delta$ and $R_{phot}$.}

\figcaption [] {$\Delta$ as a function of peak surface brightness for 429
galaxies. The dashed line is the best fit.}

\figcaption []{$\Delta$ as a function of color for
508 galaxies. The dashed line is the best-fitting relation between
$\Delta$ and $(V-R)_{CCD}$.} 

\figcaption [] {Surface brightness $vs.$ color index for the
galaxies in the OCS sample with $15.5 < R < 16.13$ mag.}

\figcaption []{$(V-R)_{CCD}$ color distributions for each morphological type. 
Each panel gives the mean color and the number of galaxies, N.}

\figcaption [] {Cone diagram for the OCS for $z \leq 0.15$.
For clarity, we omit galaxies 
with $cz \leq 1000$ km s$^{-1}$.}

\figcaption [] {Cone diagram for the DCS. Filled squares represent galaxies
in the OCS with $R_{phot} \leq 16.13$. Open circles are galaxies
with $16.13 < R_{phot} \leq 16.4$. The redshift limits are the same as
in Figure 6.}

\figcaption [] {Redshift histograms for (a) the OCS and (b) the DCS.
In the DCS, the solid histogram refers to the galaxies also in the OCS;
the dashed histogram refers to galaxies with $R_{phot} \leq 16.4$.}

\end{document}